\documentclass[aps,twocolumn,showpacs,pre,superscriptaddress,lengthcheck]{revtex4}

\usepackage[dvips]{graphics}
\usepackage{graphicx}
\usepackage{amsfonts}
\usepackage{amssymb}
\usepackage{amsmath}
\usepackage{subfigure}

\begin{document}

\title{Ground-state Properties of Small-Size Nonlinear Dynamical Lattices}
\author{P. Buonsante}
\affiliation{Dipartimento di Fisica, Politecnico di Torino, Corso Duca degli Abruzzi 24, I-10129 Torino, Italy}
\author{P. G. Kevrekidis}
\affiliation{Department of Mathematics and Statistics, 
University of Massachusetts, Amherst MA
01003-4515, USA}
\author{V. Penna}
\affiliation{Dipartimento di Fisica, Politecnico
  di Torino, Corso Duca degli Abruzzi 24, I-10129
  Torino, Italy}
\author{A. Vezzani}
\affiliation{Dipartimento di Fisica and CNR-INFM,
  Universit\`a degli Studi di Parma, Parco Area
  delle Scienze 7/a, I-43100 Parma, Italy}

\begin{abstract}
We investigate the ground state of a system of interacting
particles in small nonlinear lattices with $M \geq 3$ sites,
using as a prototypical example the discrete nonlinear Schr{\"o}dinger
equation that has been recently used extensively in the contexts of
nonlinear optics of waveguide arrays, and Bose-Einstein condensates
in optical lattices. 
We find that, in the presence of attractive interactions,
the dynamical scenario relevant to the ground state and
the lowest-energy modes of such few-site nonlinear lattices reveals
a variety of nontrivial features that are absent
in the large/infinite lattice limits:
the single-pulse solution and the uniform solution are found to
coexist in a finite range of the lattice intersite coupling 
where, depending on the latter, one of them represents the ground state; 
in addition, the single-pulse mode 
does not even exist beyond a critical parametric threshold. 
Finally, the onset of the ground state (modulational) instability appears 
to be
intimately connected with a non-standard (``double transcritical'') type
of bifurcation that, to the best of our knowledge, has not been reported
previously in other physical systems.  
\end{abstract}


\maketitle

\section{Introduction} 
\label{intro}
In the past few years, there has been a tremendous increase
in the number of studies of lattice dynamical systems, especially in the
context of differential-difference equations, where the 
evolution variable is continuum and the spatial dependence
is inherently or effectively posed on a lattice \cite{reviews}.
Such settings appear to be ubiquitous in very diverse physical
contexts ranging from the spatial dynamics of optical beams in 
coupled waveguide arrays in  nonlinear optics \cite{reviews1} to 
the dynamical behavior of Bose-Einstein condensates (BECs) 
in optical lattices in soft-condensed matter physics \cite{reviews2}, 
and even the DNA double strand in biophysics \cite{reviews3}, among others. 

One of the principal foci of this research effort 
is the analysis of the features of the localized, solitary
wave solutions of such lattices. Discrete solitons 
\cite{solit}, and various more exotic structures such 
as dipole solitons, soliton-trains, soliton-necklaces 
and vector solitons were recently observed in optical
contexts such as photorefractive materials \cite{esolit}.
At the same time, experimental developments in the physics of   
BECs closely follow with prominent recent results, including the 
observation of bright, dark and gap solitons in quasi-one-dimensional 
settings \cite{bec}.

Another trend that has recently been followed is to study
small lattices, such as those pertaining to double or 
triple-well potentials. The aim there is to better understand the
underlying physics of such simpler dynamical systems and
subsequently explore how much of the relevant phenomenology
may persist in the infinite lattice limit.
It is interesting to note that  few-site lattices were among the first
ones to be explored thoroughly, starting with the pioneering work 
of \cite{eilbeck}.  Since then, a variety of theoretical
works also examined features relevant especially to double-well 
(such as symmetry-breaking \cite{weinstein}), triple-well
(such as oscillatory instabilities \cite{johansson} and 
chaotic behavior \cite{penna}, among others), or even multi-
(but few-) well potentials (such as synchronization \cite{pando}).
Most of the above studies were done in the prototypical 
nonlinear envelope wave equation that is equally applicable
to each of the above mentioned physical settings (in appropriate
parameter regimes), namely the discrete nonlinear Schr{\"o}dinger
equation (DNLS).

In that vein, recent experiments in both optical media \cite{zhigang} and
in BECs \cite{markus2} have revealed a host of interesting 
phenomena such as symmetry breaking in double-well potentials,
and constructive (destructive) interference of in (out-of-) phase
pulses in triple-well media, among others. 
More generally, few-site lattices such as those we are going
to address in what follows appear to be within the reach of state-of-art
technology in optically trapped BECs. Actually, Ref. \cite{A:Albiez} reports
on the analysis of the evolution of the density distribution and
relative phase of a boson Josephson junction. Such a two-site system  was
realized by isolating a single ``edge'' of an optical lattice via
an additional confining potential. 
Likewise,  single ``plaquettes'' of suitable two-dimensional --- possibly
quasiperiodic --- lattices  \cite{rings1} can be used to create few-site
closed rings. A further interesting proposal is based on transverse
electromagnetic modes of laser beams \cite{rings2}.

Our main focus here is on the ground-state properties
of a system of interacting particles, hopping in a few-site lattice,
described by the DNLS equation.
We emphasize that this is an important issue for
any low- (virtually zero-) temperature system,
in particular for ultracold bosons. In this respect, on lattices, 
the mean-field description in terms of the DNLS equations,
which results from a variational approach to the quantum ground-state,
proved to be quite satisfactory with both repulsive \cite{ampe} and
attractive interactions \cite{praBPV,jack}.
Naturally, a great deal of the relevant phenomenology
has been analyzed. However, we hope to illustrate  
that there some of the properties 
of low energy states are still unexplored and yet are particularly intriguing
in simple and interesting systems such as the few-site lattices. 
As  is well known, in the case of attractive, ``focusing''
interaction, the ground state of the homogeneous systems under
investigation exhibits a delocalization transition \cite{reviews}
driven by the effective intersite coupling $\epsilon$.  
The threshold for localization is in general identified with the
occurrence of modulational instability in the uniform ground state
characterizing the system at large values of $\epsilon$.
Here we show that this identification applies only to sufficiently
large lattices. Conversely, on small lattices, these two thresholds
are distinct, the delocalization transition occurring at a larger value
of $\epsilon$.
As we will illustrate, this feature can be
understood in terms of the complex interplay of three low-energy solutions
to the DNLS equations governing the system.  These are the uniform state
and two localized solutions that, for reasons that will become clear shortly,
we refer to as   {\it single-pulse}  and  {\it two-pulse} state, respectively.
These two localized solutions emerge as excited states from a saddle-node
bifurcation point below some threshold in $\epsilon$.
As this parameter is further lowered below the delocalization threshold, the
symmetry-breaking single-pulse state becomes the ground state of the
system, but this does not influence the stability properties of the
uniform state. 
As we will show, this feature can be in principle
exploited to access metastable states of the system.
Lowering 
$\epsilon$ further eventually results in the
uniform-state modulational instability, 
caused by a bifurcation involving the uniform state and the two-pulse state.
As we discuss in Section \ref{S:numres}, this critical point, which
we dub {\it double transcritical bifurcation}, exhibits non-standard
features which, to the best of our knowledge, have not been observed
previously. 

The layout of the paper is the following. In section \ref{S:setup}
we introduce the model and recall the known results about its ground
state. In Section \ref{S:pert} we show that interesting insight in the
bifurcations involving the uniform state can be gained through a simple
perturbative approach. In Section \ref{S:trimer} this analytical 
insight is fully developed for the case of the three-site lattice. 
Section \ref{S:numres} contains 
the numerical results for lattices comprising of 
$M=3,4,5$ sites. In particular,
we discuss the non trivial features of the double transcritical bifurcation
related to the uniform-state modulational instability, and illustrate
how the known situation for large lattices is recovered for $M \geq 6$.
Our conclusions are given in Sec. \ref{S:concl}.

\section{Setup}
\label{S:setup}
In the following, we consider the standard DNLS equation on a $M$-site one-dimensional closed lattice,
\begin{equation}
\label{dnlse1}
i \dot z_n=-T \Delta_2 z_n - \Gamma |z_n|^2 z_n,
\end{equation}
where $\Delta_2 z_n=(z_{n+1}+z_{n-1}-2z_n)$ is the discrete Laplacian, and
periodic boundary conditions are implemented by identifying sites $n=1$ and
$n=M+1$. This equation is derived from the Hamiltonian
\begin{eqnarray}
\label{E:H1}
H=\sum_{n=1}^M T |z_{n+1}-z_n|^2 - \frac{\Gamma }{2} |z_n|^4
\end{eqnarray}
making use of the standard Poisson brackets $\{z_n, z_m\} = i \delta_{n m}$ \cite{reviews}.  We recall that Hamiltonian (\ref{E:H1}) is the semiclassical counterpart of the Bose-Hubbard model  standardly adopted for describing ultracold bosonic atoms trapped in optical lattices \cite{fisher,jaksch}. In more detail, Eq. (\ref{E:H1}) is obtained by approximating the quantum states with suitable coherent states and subsequently implementing a standard variational method.
Direct comparison shows that  Hamiltonian (\ref{E:H1}) provides a description of the ground state properties of the fully quantum model that is  satisfactory in many respects, both in the case of repulsive \cite{ampe} and attractive interactions \cite{praBPV}. We also recall that in this framework  $T$ and $\Gamma$ --- denoting the intersite coupling across adjacent sites and the boson-boson interaction, respectively --- are directly related to experimental parameters that can be varied over a wide range of values \cite{jaksch,theis}. As to $z_n$, it is a macroscopic complex variable describing the bosonic population, $|z_n|^2$, and phase, $\arg(z_n)$, at lattice site $n$. It is easy to prove that the  total population $N = \sum_{n=1}^M |z_n|^2$ is  conserved along the dynamics \cite{note1}.

As we mention in Sec. \ref{intro}, we focus on the case of 
attractive interactions, $\Gamma > 0$. 
 Before proceeding with our discussion, we observe that the only independent parameter in Eq. (\ref{dnlse1}) other than the lattice size $M$ is the effective (rescaled) intersite coupling, $\epsilon = T/(\Gamma N)$. Actually,  Eq. (\ref{dnlse1}) can be recast in the form
\begin{eqnarray}
i \dot{u}_n=-\epsilon \Delta_2 u_n - |u_n|^2 u_n,
\label{dnls2}
\end{eqnarray}
where $u_n = z_n/\sqrt N$ and the dot now denotes 
the derivative with respect to the rescaled time variable $t' = \Gamma N t$.

We are interested in the ground state of the Hamiltonian (\ref{E:H1}),
i.e. in the state $v_n$ minimizing
\begin{equation}
\label{E:Rene}
E= \frac{H}{\Gamma N^2} = \sum_{n=1}^M \epsilon |v_{n+1}-v_n|^2 -
\frac{1}{2} |v_n|^4
\end{equation}
Therefore $v_n$ satisfies the equation
\begin{eqnarray}
G(v_n;\epsilon)=\Lambda v_n- \epsilon \Delta_2 v_n - |v_n|^2 v_n=0.
\label{dnls3}
\end{eqnarray}
where the eigenvalue $\Lambda$ is a Lagrange multiplier taking into account the constraint $\sum_n |v_n|^2 = \sum_n |u_n|^2 = 1$ stemming from the norm conservation.
We note that the solutions to the nonlinear eigenvalue problem Eq. (\ref{dnls3}) correspond to the standing wave solutions to Eq. (\ref{dnls2}) of the 
form $u_n = v_n e^{i\Lambda t'}$.

Let us now recall some well known facts about this ground state.
For small values of the intersite coupling $\epsilon$, the ground state of the system is known to break the translational invariance of $H$. Actually for  vanishing $\epsilon$'s, i.e.,  in the so-called {\it anticontinuum limit}, it is easy to check that the ground state is completely localized at a single lattice site $n_0$, $u_n = \delta_{n\, n_0}$. As the intersite coupling is increased, the width of the localization peak increases while maintaining its single-pulse profile, i.e.  remaining mirror-symmetric with respect to the central site $n_0$, $u_{n_0+k} = u_{n_0-k}$. Hence we refer to this solution of Eq.  (\ref{dnls3}) as {\it single-pulse} state. Note that the localization peak of the single pulse state can be centered at any lattice site, so that the symmetry breaking ground-state is $M$-fold degenerate.

Conversely, for sufficiently large values of $\epsilon$, the translational symmetry is recovered, the ground state being  the uniform (i.e. delocalized) state  $v_n = v = M^{-1/2}$, of energy $E = 1/(2M)$. 
The (finite) critical value of the intersite coupling at which the inversion in the nature of the ground state occurs is referred to as delocalization threshold. This critical value is usually identified with the  threshold below which the uniform state becomes modulationally unstable \cite{jack},
\begin{eqnarray}
\epsilon_1(M)
= \frac{1}{2M} \frac{1}{\sin^2\left(\frac{\pi}{M}\right)}.
\label{dnls10}
\end{eqnarray}
In the following we will show that this identification is correct only for $M\geq 6$, whereas on smaller lattices the delocalization threshold occurs at a critical value $\epsilon_2 > \epsilon_1(M)$. Furthermore we will show that the threshold for modulational instability corresponds to a non-standard bifurcation involving the uniform state and a low-energy localized solution to  Eq.  (\ref{dnls3}). We will refer to the latter as {\it two-pulse state} since, unlike the single-pulse state, its localization peak features a maximum at
two adjacent sites, reducing to  $u_n = (\delta_{n\, m} + \delta_{n\, m+1})/\sqrt 2$ in the anticontinuum limit.

\section{Perturbative Approach}
\label{S:pert}
%
In this section we focus on the possible bifurcations involving the
uniform state $v_n = 1/\sqrt M$, which is the ground-state of the
system for sufficiently large $\epsilon$'s.  Hence we look for
nonuniform solutions to Eq. (\ref{dnls3}) that become uniform as
the intersite coupling $\epsilon$ approaches a finite value.
Adopting a simple perturbative approach, we introduce a linear parameter
$\tau$ such that $\epsilon = \alpha_\epsilon + \beta_\epsilon \tau$
and assume that states of the form $v_n =  v + \tau (p_n + i q_n)$
satisfy Eq. (\ref{dnls3}) with $\Lambda = v^2 + \beta_\Lambda \tau$.
After some simple manipulations one gets
\begin{eqnarray}
\label{lin1}
\beta_\Lambda v - \alpha_\epsilon \Delta_2 p_n  - 2 v^2 p_n &=& 0 \\
\label{lin2}
 \alpha_\epsilon \Delta_2 q_n &=& 0
\end{eqnarray}
According to Eq. (\ref{lin2}), the imaginary part of the perturbation
is uniform, $q_n = q$. Hence,  it can be absorbed in the unperturbed
uniform state as a phase factor, $v \to |v| e^{i \theta}$, with
$\theta = \arcsin (\tau q/|v|)$.
As to the real part, it is easy to prove that the coefficient $\beta_\Lambda$
appearing in Eq. (\ref{lin1}) must vanish, which suggests that spatially
modulated solutions branch-off tangentially from the uniform state.
This is obtained summing  Eq. (\ref{lin1}) over $n$ and making use of the
constraint $\sum_n p_n = 0$ stemming at linear order from the normalization
for the perturbed solution, $\sum_n |v_n|^2 = 1$. Hence Eq. (\ref{lin1})
is formally equivalent to the eigenvalue equation for the discrete
Laplacian operator 
\begin{equation}
\Delta_2 p_n =  \lambda p_n, \qquad  \lambda = - \frac{2 v^2}{\alpha_\epsilon} 
= - \frac{2}{M \alpha_\epsilon}
\end{equation}
On a $M$-site homogeneous  one-dimensional lattice such as that under
investigation,  the Laplacian features $M$ eigenvalues of the form
$\lambda_k = -4 \sin^2 (\pi k/ M)$, with $k = 0,\ldots,M-1$. These
define a set of  critical values for the intersite coupling,
$\alpha_\epsilon^{(k)}=[2 M \sin^2 (\pi k/ M)]^{-1}$, where
nonuniform solutions become uniform. 
The relevant perturbative modulations have the form
$p_n^{(k)} \propto \sin ( 2\pi k n/M + \varphi)$, where
$\varphi$ is a phase that ensures that two solutions corresponding
to the same $k$ are  independent.
Note that $k=0$ must be discarded, since it corresponds to a
vanishing perturbation, and that
$\alpha_\epsilon^{(k)} = \alpha_\epsilon^{(-k)} = \alpha_\epsilon^{(M-k)}$.
Hence, according to this picture, one expects $\lfloor (M-1)/2 \rfloor$
distinct critical values, where $\lfloor x \rfloor$ denotes the largest
integer smaller than $x$. 

We now observe that $\alpha_\epsilon^{(k+1)} < \alpha_\epsilon^{(k)}$,
and that the largest critical value $\alpha_\epsilon^{(1)}$ coincides
with the known threshold for modulational instability reported in
Eq. (\ref{dnls10}).
This means that modulational instability occurs in correspondence to
a bifurcation point where nonuniform solutions merge with the uniform
state. Note that, for suitable choice of the phase $\varphi$, these
nonuniform solutions $u_n = v + \tau p^{(1)}_n$ may have either a 
single-pulse or a two-pulse character.
Explicit analytic results for the three-site lattice and numeric results
for larger lattices, reported respectively in Secs. \ref{S:trimer} and
\ref{S:numres}, will confirm this scenario. These results will also
evidence the non-trivial character of the bifurcation point corresponding
to the onset of modulational instability.
As to the remaining critical points, it can be proved that they correspond
to bifurcations involving nonuniform solutions that in the anticontinuum
limit $\epsilon \to 0$ reduce to the form
$u_n = P^{-1/2} \sum_{r=1}^{P} \delta_{n\, n_r}$,
where  $n_r$ denotes different lattice sites. Since these bifurcations
occur when the uniform state is unstable, and therefore not the ground
state of the system, their detailed study goes beyond the purpose of
this paper.

\section{Analytical results for $M=3$}
\label{S:trimer}
We now turn to the analytically tractable case of the three-site lattice.
As illustrated in the previous sections, the uniform,  single- and two-pulse solutions to Eq. (\ref{dnls3}) are expected to play a significant role in relation to the ground state of the system. In this simple case, all of these three states correspond to a triplet 
$(v_1, v_2, v_3)$ with $v_1 = v_3$. Hence they can be described by a 
unique parameter $v = v_1/v_2$. Clearly $v$ equals 1 for the uniform state, whereas it varies in the intervals $[0,1]$ and $[1,\infty]$ for the single- and two-pulse state, respectively.

\begin{figure}
\centerline{
\includegraphics[width=4.cm,height=5cm,angle=0,clip]{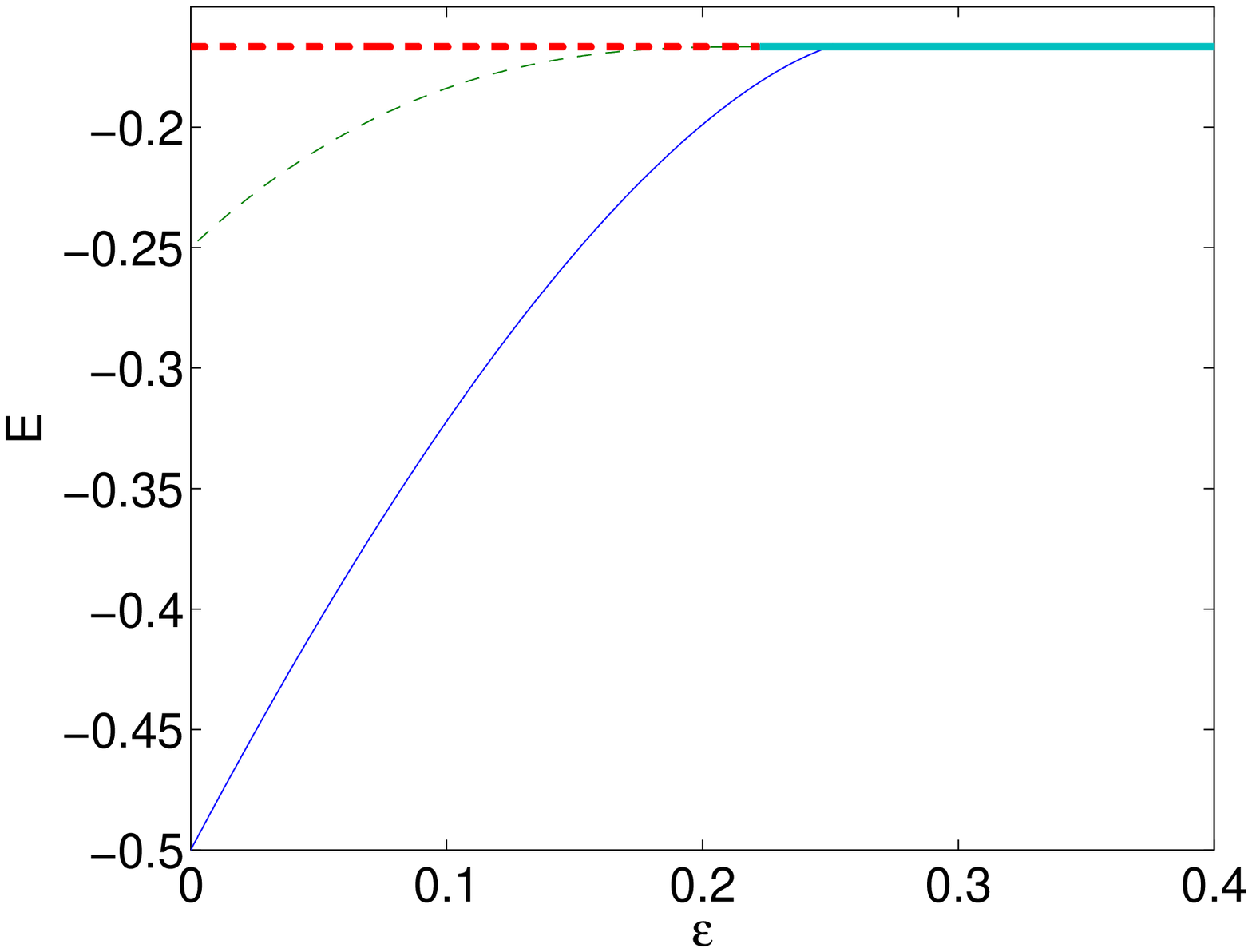}
\includegraphics[width=4.cm,height=5cm,angle=0,clip]{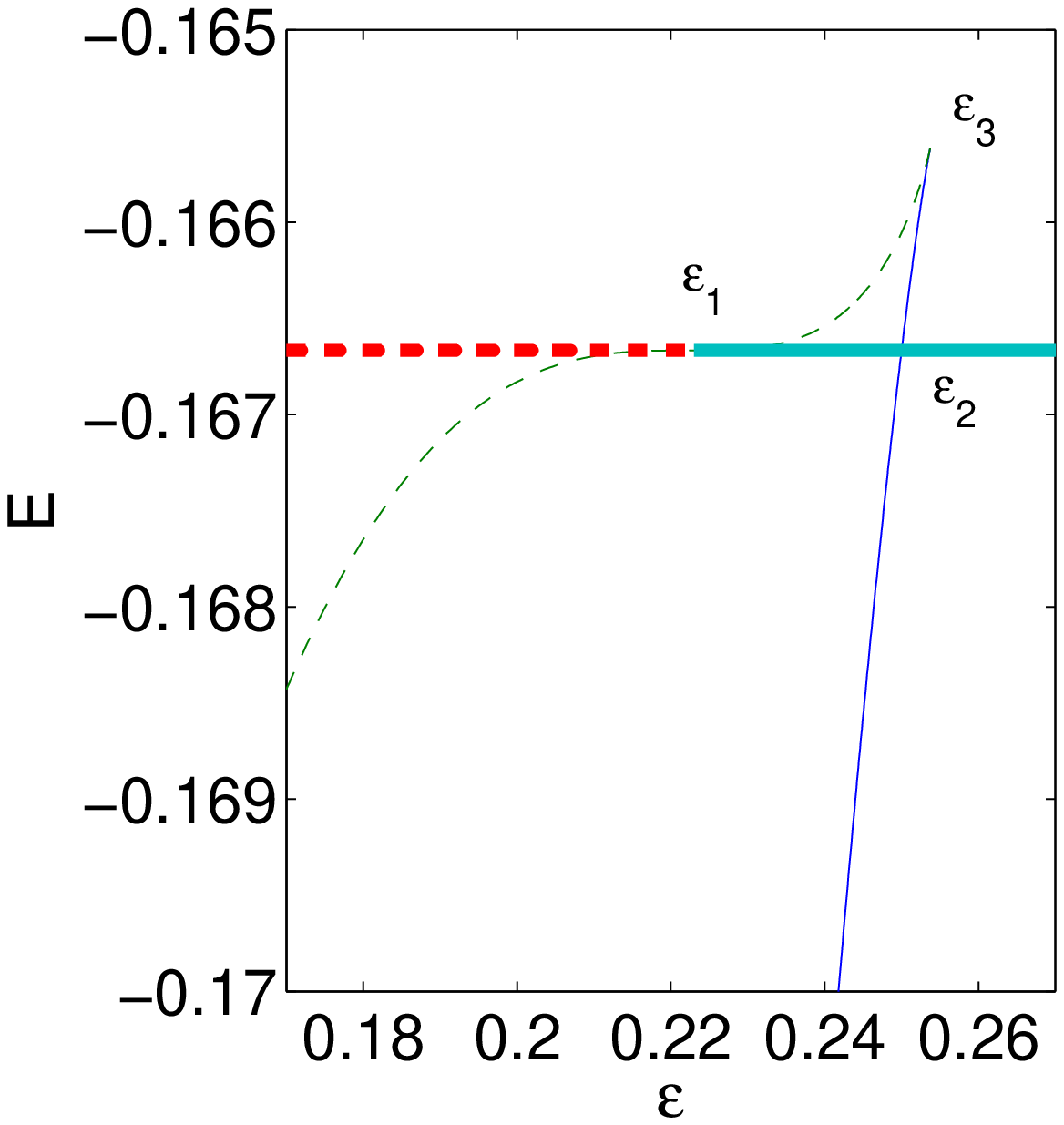}
}
\caption
{Energy $E$ of the branches considered as
a function of $\epsilon$ for a three-site lattice, as
provided by Eq. (\ref{trimer}) and confirmed by the numeric analysis
in Sec. \ref{S:numres}.
Thin solid: stable single pulse; thin dashed: unstable
two-pulse (becomes single-pulse-like for $\epsilon>\epsilon_1$);
thick: unstable (dashed) and stable (solid) portions of
the uniform branch. The right panel is a blowup of the left one, 
and clearly illustrates the relevant critical points.
}
\label{dfig0}
\end{figure}

Plugging this form into Eqs. (\ref{dnls3}) and (\ref{E:Rene}) and making use of the normalization constraint, after some algebra one gets the parametric description
\begin{equation}
\label{trimer}
\begin{array}{cc}
{\displaystyle \epsilon=\frac{v(1+v)}{(1+2v)(1+2v^2)}}, \\
\\
{\displaystyle E = -\frac{1-2v +4v^2+4v^3-2v^4+4v^5}{2(1+2v)(1+2v^2)^2},}
\end{array}
\end{equation}
Note that for $v=1$, i.e. when the two-site branch
meets the uniform branch, the first of Eqs. (\ref{trimer}) coincides
with Eq. (\ref{dnls10}) describing the critical threshold for
modulational instability, i.e. $\epsilon = \epsilon_1(3) = 2/9$.
Of course in this situation the two solutions (uniform and 
two-site) have the same
energy  $E=-1/6$. The parametric function in
(\ref{trimer}) crosses the same value of the energy at a second
point, $\epsilon=\epsilon_2 = 0.25 > \epsilon_1(3)$, corresponding to
a single-pulse solution. 
Actually both of the functions in Eq. (\ref{trimer}) feature a maximum
at the same value of $v$, corresponding to
$\epsilon=\epsilon_3 \approx 0.2537 > \epsilon_2$,
while $v\to \infty$ corresponds to $\epsilon=0$ and $E= -0.25$.
This means that for $\epsilon_1<\epsilon<\epsilon_3$ there are,
in fact, two single pulse branches with different energies
for a given $\epsilon$. 
The most energetic of them emerges from the two-pulse (and
uniform) branch at $\epsilon_1$, whereas the least energetic exists 
also in the interval $[0,\, \epsilon_1]$, where it is the 
ground state of the system.
As to the stability properties, it can be shown analytically that the
low-energy single-pulse branch is always stable, while the high-energy
single-pulse and the two-pulse branch are always unstable. As mentioned
above, the uniform branch is unstable below and stable above $\epsilon=
\epsilon_1(3)$. The situation for $M=3$ is corroborated
by numerical bifurcation results (that are detailed below) in Fig. \ref{dfig0}.
Interestingly, the right panel highlights the presence of an inversion in the
nature of the ground state at $\epsilon =\epsilon_2$ which appears to
coincide with the single-pulse solution (uniform solution)
for $\epsilon <\epsilon_2$ ($\epsilon >\epsilon_2$).

%
%
%
\begin{figure}
\centerline{
\includegraphics[width=4.cm,height=5cm,angle=0,clip]{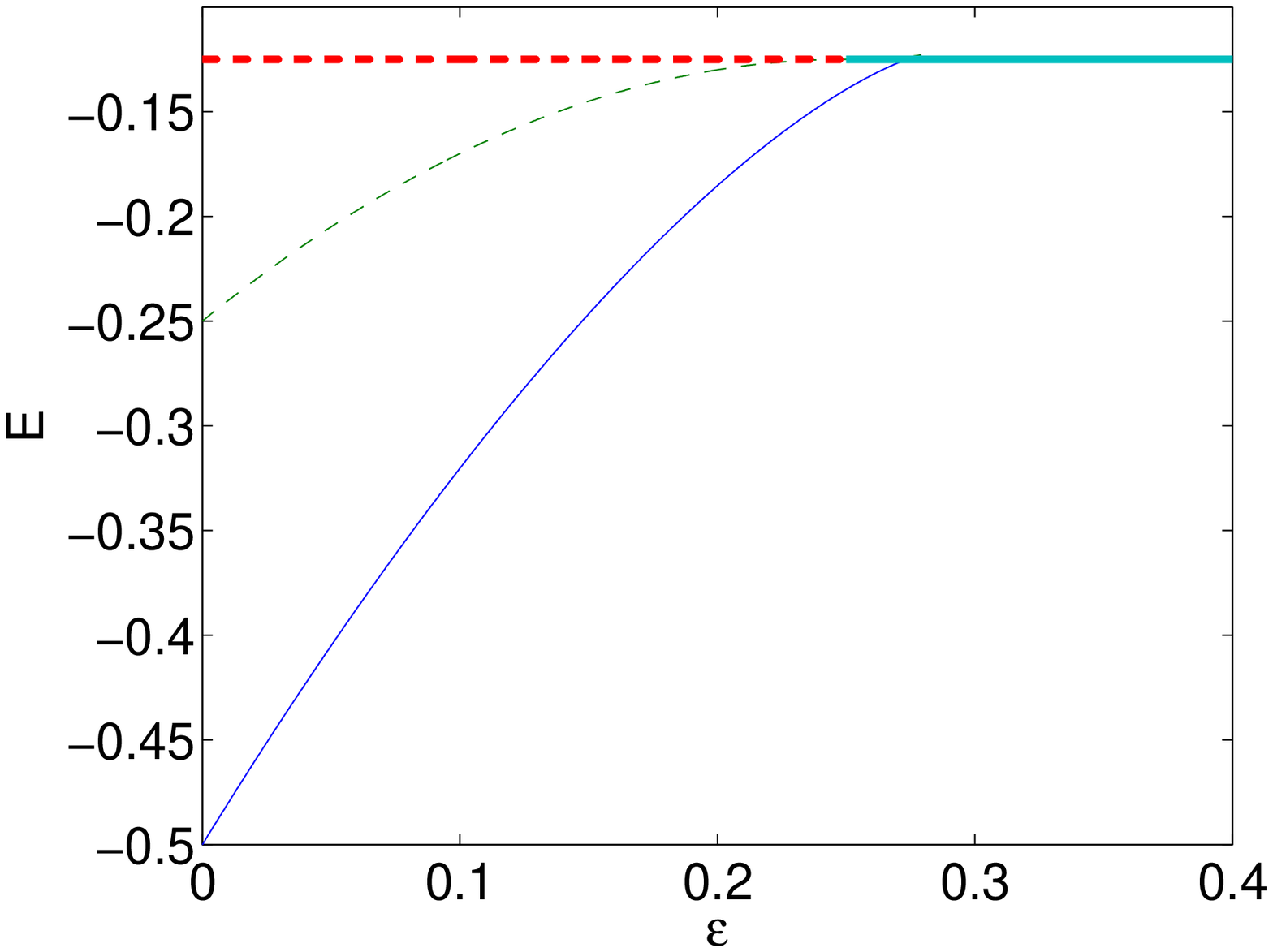}
\includegraphics[width=4.cm,height=5cm,angle=0,clip]{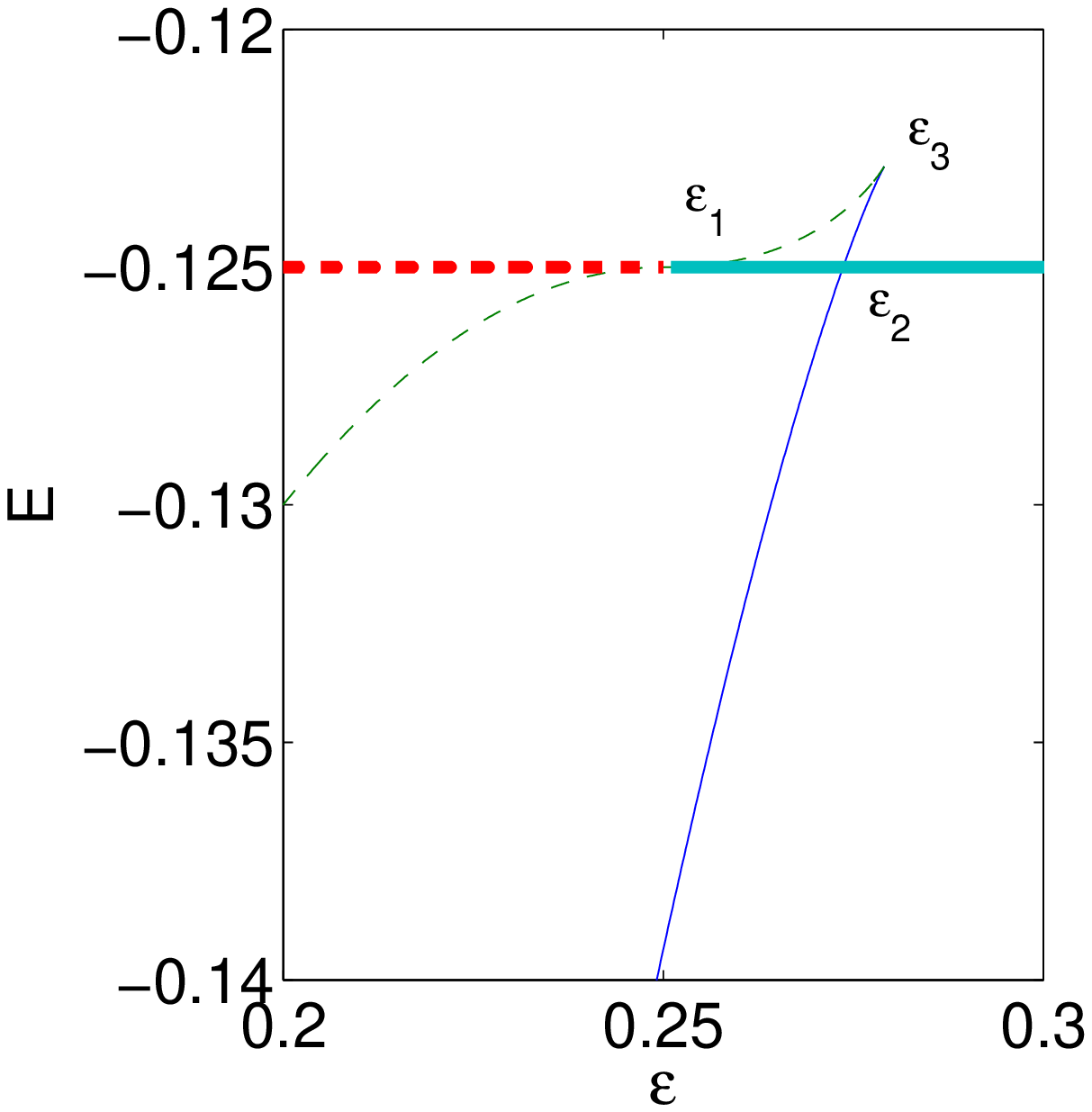}
}
\caption
{
Energy $E$ of the solution branches considered as
a function of $\epsilon$ for a four-site lattice, as resulting
from the numeric analysis in Sec. \ref{S:numres}.
The line styles and colors have the same meaning as in Fig. \ref{dfig0}.
The right panel is a blowup of the left one, 
and clearly illustrate the relevant critical points.
}
\label{dfig1}
\end{figure}
%

\section{Numerical techniques and results}
\label{S:numres}

A numerical study of the single- and two-pulse solution can be
efficiently performed by means of Keller's pseudo-arclength continuation method \cite{keller}.
This allows us to trace the relevant branches of solutions past fold 
points. Given a solution $({v_n}^{(0)}, \epsilon^{(0)})$ of the equation
$G({v_n};\epsilon)=0$ 
and a `direction' vector $({{\bar v}_n}^{(0)},{\bar \epsilon}^{(0)})$,
one can derive $({v_n^{(1)}},\epsilon^{(1)})$ by solving the system of equations
\begin{equation}
\begin{array}{c}
G_1 \equiv G({v_n^{(1)}},\epsilon^{(1)})=0,\\
\\
({v_n^{(1)}}-{v_n^{(0)}})*{\bar v}_n^{(0)}
+(\epsilon^{(1)}-\epsilon^{(0)})
{\bar \epsilon}^{(0)}-\Delta s=0,
\end{array}
\label{5}
\end{equation}
where $\Delta s$ is a pre-selected arclength parameter (we typically used
$\Delta s=0.001$). The parenthetic superscript denotes the iteration step
index.
Subsequently, one can use Newton's method to solve the system in equation
\ (\ref{5}).
The next (normalized) `direction' vector 
$( {\bar v}_n^{(1)} ,{\bar \epsilon }^{(1)})$,
is then computed by solving
\begin{equation}
\left(
\begin{array}{cc}
\frac{\partial}{\partial v_n} G_1 & \frac{\partial}{\partial \epsilon} G_1 
\\ 
{\bar v}_n^{(0)} & {\bar \epsilon}^{(0)} 
\end{array}
\right)
\left(
\begin{array}{c}
{\bar v}_n^{(1)} \\ {\bar \epsilon}^{(1)}
\end{array}
\right) =
\left(
\begin{array}{c}
0 \\ 1
\end{array}
\right),
\label{7}
\end{equation}
and the process is then iterated. In this setting, there is a natural
starting point of this iteration process at $\epsilon=0$, where the
equation (\ref{dnls3}) becomes algebraic. In that limit, the  ``single pulse''
branch is given by $v_n=\delta_{n,n_0}$, with support over the site $n_0$,
and the ``two-site'' pulse by $v_n=(\delta_{n,n_0}+\delta_{n,n_0+1})/\sqrt{2}$.
These branches are initialized
with the above exact profile in this anti-continuum limit of $\epsilon=0$,
and subsequent continuation of the solutions allows their path-following,
as the parameter $\epsilon$ is varied. For each step, once these solutions
are obtained, their numerical linear stability is performed by 
using:
\begin{equation}
u_n=e^{i\Lambda t} \left[v_n+\delta \left( a_n e^{- \lambda t}+
b_n e^{\lambda^{\ast} t}\right) \right],
\label{8}
\end{equation}
where $\delta$ is a formal linearization parameter.
This results into a linear (matrix) eigenvalue problem 
for $(\lambda,\{a_n,b_n^{\star}\})$ that we also
solve. 

\begin{figure}
\centerline{
\includegraphics[width=4.cm,height=5cm,angle=0,clip]{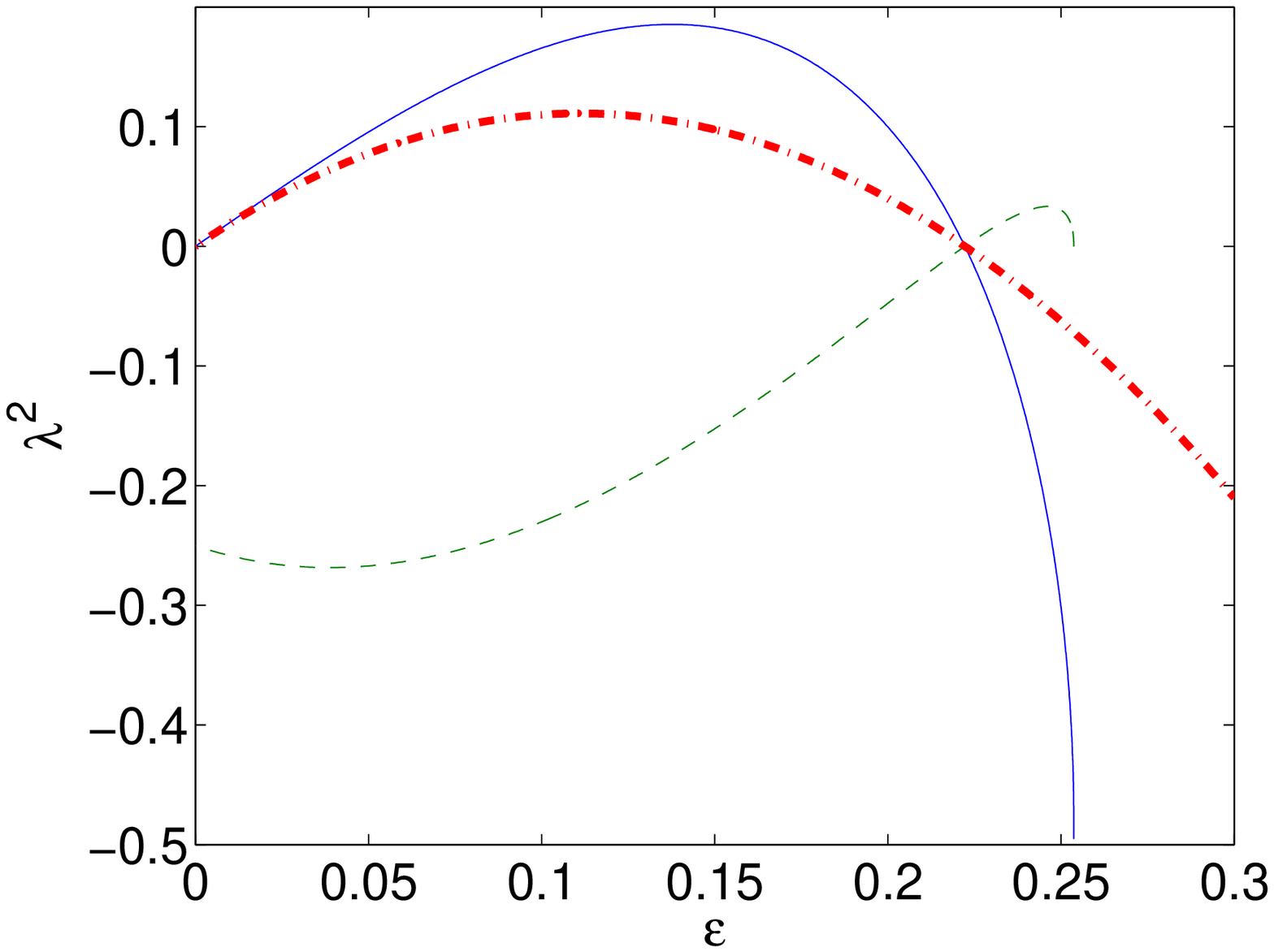}
\includegraphics[width=4.cm,height=5cm,angle=0,clip]{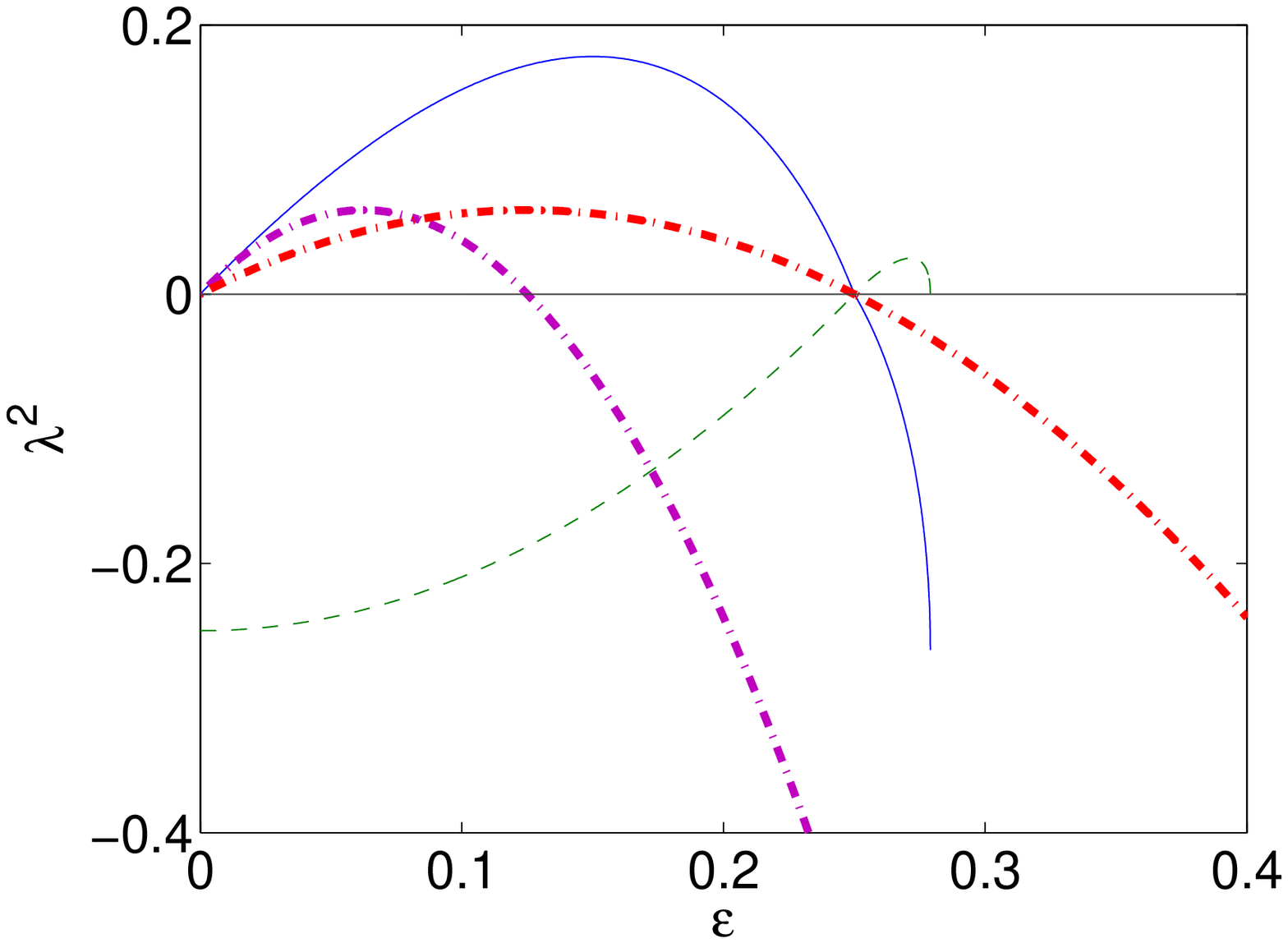}
}
\caption{
Squared eigenvalues $\lambda^2$ of the $M=3$ (left) and $M=4$ (right)
case, as a function of $\epsilon$. Thick (thin) lines refer to the
uniform (two-site pulse) branch. Note that for $M=4$ an additional
 eigenvalue for the uniform state exists and is shown.} 
\label{dfigB}
\end{figure}

\begin{figure}
\centerline{
\includegraphics[width=4.cm,height=5cm,angle=0,clip]{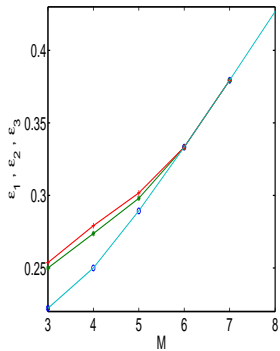}
}
\caption{
Location of the critical points discussed in the text, with increasing lattice size $M$. Circles, stars and plus symbols denote
$\epsilon_1$, $\epsilon_2$ and $\epsilon_3$, respectively. The line joining
the circles shows the theoretical prediction of Eq. (\ref{dnls10}). The other lines are mere guides to the eye.}
\label{dfig3}
\end{figure}

The analytical description obtained above for the $M=3$ case is confirmed by our numerical results, and remains qualitatively the same for $M=4$ and $M=5$. The situation for $M= 4$ is illustrated in figure \ref{dfig1} and \ref{dfigB}.
%
%
Figure \ref{dfig1} shows the behaviour of the energy as
a function of $\epsilon$. As in Fig. \ref{dfig0}, the right
panel is a zoom into the most interesting region. Solid lines
denote stable branches, whereas dashed lines denote unstable
branches, as specified in the captions. The evident analogies
between the three- and four-site lattices
(also present for $M=5$, not shown) can be summarized as follows.
At the critical point where the modulational instability
arises, $\epsilon=\epsilon_1(M)$, the 
two-pulse branch ``collides'' with the uniform branch, and ``emerges''
from it as a (higher-energy) single site-branch. This  eventually collides
with the low-energy single-site branch at $\epsilon=\epsilon_3(M)$.
The latter originates from the single-pulse solution at $\epsilon=0$,
and is the ground state of the system until it  crosses the uniform 
branch at $\epsilon=\epsilon_2(M)<\epsilon_3(M)$.
Note however that this crossing is not a collision in the bifurcation sense,
since the configurations of the two branches remain different at this point. 
They merely have the same energy for fixed norm.

Also, the collision occurring at $\epsilon_1(M)$ appears to be definitely
non-standard, from a bifurcation theory point of view. 
This is not only because of the ``tangency''
%
%
of the two branches at the critical point, but also due to the
fact that, contrary to what would be expected from such an 
apparent transcritical bifurcation, the branches do not exchange
their stability, but rather the 
two-pulse branch {\it remains} linearly unstable (before, as well as
after the collision). 
Further insight in the non-standard nature of the bifurcation
at $\epsilon_1(M)$ is gained from Fig. \ref{dfigB}, showing,
for both $M=3$ and $M=4$, the
crucial squared eigenvalues of the two-site and uniform branches, as
resulting from our numerical analysis.
More specifically, the principal (i.e., maximal)
eigenvalue responsible for the instability of the uniform mode
turns out to be a {\it double} eigenvalue. 
This double eigenvalue approaches $\lambda=0$, as $\epsilon \rightarrow
\epsilon_1$ (recall that stabilization implies that this
real eigenvalue pair should become imaginary as $\epsilon$ crosses
$\epsilon_1$, hence its square should change sign). For
the two-site branch, an imaginary eigenvalue (for $\epsilon<\epsilon_1$, 
shown by green line in the figures)
tends to zero (and becomes real for $\epsilon>\epsilon_1$). 
However, in order for the multiplicity to be preserved (given the
double eigenvalue of the uniform state), an additional eigenvalue
should cross zero at this critical point (this time, coming from
the real side, namely the blue line in Fig. \ref{dfigB}). As a result,
along the former eigendirection, indeed there is a transcritical exchange
of stability, however, the latter eigendirection enforces an additional
change of stability for the two-site branch.
This results in a novel and non-standard scenario that we call the
``double transcritical'' bifurcation resulting in one of the branches
being {\it unstable before and unstable after} the critical point.


Concerning the ground-state properties of such small-size lattices,
the above analysis and numerics confirm an important feature made
visible by the analytical study of the case $M=3$, 
that is the presence of a critical
value of $\epsilon$ where the inversion
in the nature of the ground state takes place.  
This feature, in fact, does {\it not always} occur at the 
critical point for the modulational instability of the uniform state,
$\epsilon_1$, but rather at the crossing point $\epsilon_2$ previously
discussed.
That is to say, there is an interval $I_1 = [\epsilon_1,\epsilon_2]$
where the ground state is localized despite that the uniform state is
modulationally stable. Likewise, there is an interval
$I_2 = [\epsilon_2,\epsilon_3]$ where the ground state is uniform, and the
single-pulse state is an excited (i.e., higher energy for the same norm)
stable state. A further feature worth emphasizing is that the latter
terminates at a finite value, $\epsilon=\epsilon_3$, due to its collision
with the high-energy single-pulse branch discussed above. 
%
This is perhaps contrary to the common intuition based on the infinite 
lattice \cite{miw}, where the single-pulse branch exists up to the continuum
limit, $\epsilon\to \infty$.
%

Finally, Fig. \ref{dfig3} shows the location of the critical points 
$\epsilon_1$, $\epsilon_2$ and $\epsilon_3$ with increasing lattice sizes 
$M$. This indicates how to reconcile the above picture with 
the infinite lattice limit \cite{kapitula}, where the uniform state is always 
modulationally unstable, the single-pulse state is always the ground state, 
and the latter collides with the two-pulse branch only at $\epsilon 
\rightarrow \infty$. More specifically, Fig. \ref{dfig3} shows that the 
picture offered above with the relevant regimes persists for 3-, 4- and 
5-site lattices, while for lattices with 6 or more sites the three critical 
points marking the boundaries of intervals $I_1$ and $I_2$ collapse to the 
single value $\epsilon_1(M)$ described by Eq. (\ref{dnls10}). That is to 
say, for $M \geq 6$ the two intervals shrink to a single point, and the 
change in the nature of the ground state occurs at the critical point 
for the modulational instability of the uniform state. Furthermore, for 
sufficiently large $M$'s, the latter is basically linear in the lattice 
size $\epsilon_1(M) \approx M/2\pi$, so that the above discussed
picture is recovered in the thermodynamic 
limit $M \to \infty$.

\section{Conclusions} 
\label{S:concl}
We have illustrated that dynamical lattices (and, in particular, small ones)
still harbor a variety of surprises.
They can host previously unraveled bifurcations (such as the 
``double transcritical'' one elucidated above); they may feature
ground state inversions, as well as  coexistence of stability between uniform 
and localized states. They may even be unable to sustain 
localized solutions for sufficiently strong tunneling. All these
numerically observed traits can also be captured analytically. 

Among the various features we have discussed in section \ref{S:numres}, 
the inversion effect characterizing the ground state of small lattices appears
to be particularly interesting.
For $M=\, 3, 4, 5$ both the uniform state and the single-pulse state are
stable solutions
of the model within the interval $\epsilon_1 \le \epsilon \le \epsilon_3$.
Such an interval is absent for $M \ge 6$.
The unexpected feature that we have evidenced is that at the intermediate value
$\epsilon_2$ of such an interval a change in the nature of the ground
state between the uniform and the single-pulse state takes place.
This inversion is driven by the parameter $\epsilon$. 
An interesting consequence
of this feature is that, at least in principle, by adiabatically
decreasing $\epsilon$ across $\epsilon_2$ the system can remain in the
uniform state without decaying in the proper (single-pulse) ground state.
A similar effect can be enacted when adiabatically increasing $\epsilon$
over $\epsilon_2$. In this case the single-pulse state (the ground state
for $\epsilon < \epsilon_2$) survives for $\epsilon > \epsilon_2$ once
more leaving the system in an excited state.
%

As discussed above, current experimental technology furnishing two-site
lattices \cite{markus2}
makes forthcoming the realization of $M>2$ small lattices.
Given the experimental tractability of
both 
optical waveguide arrays \cite{reviews1} 
and 
BECs in optical lattices \cite{reviews2} ,
the features we have shown to distinguish few-site lattices 
should have directly measurable implications in nonlinear
optics, as well as soft condensed matter physics. They also generate 
further intriguing questions, such as e.g. the origin of the
``criticality'' of the 6-site lattice which are particularly
worthwhile to address in future studies.


\medskip
{\bf Acknowledgments}. 
%
One of the authors (PB) acknowledges a grant from {\it Lagrange}
{\it project}-CRT Foundation and the hospitality of the Ultra Cold 
Atoms group at the University of Otago. PGK gratefully acknowledges
the support of NSF through the grants DMS-0204585, DMS-CAREER 
and DMS-0505663.


\begin{thebibliography}{99}
\bibitem{note1} Notice that Hamiltonian (\ref{E:H1}) differs from those in Refs. \cite{ampe,praBPV} by a term that is proportional to the total population,  $2 T \sum_n |z_n|^2 = 2 T N$. Since $N$ is conserved,  this additional term affects the dynamics of $z_n$ only through a site-independent phase factor.
\end{thebibliography}
\end{document}